\documentclass[aps,prd,twocolumn,superscriptaddress,showpacs]{revtex4-1}

\usepackage{amsmath}
\usepackage{amsfonts}
\usepackage{amssymb}
\usepackage{txfonts}
\usepackage{bm}
\usepackage{graphicx}
\usepackage{color}
\usepackage[colorlinks]{hyperref}

\definecolor{CiteColor}{rgb}{0,0.5,0}
\hypersetup{citecolor=CiteColor}
\definecolor{RefColor}{rgb}{0.55,0,0}
\hypersetup{linkcolor=RefColor}
\definecolor{darkgreen}{rgb}{0.2,0.7,0.2}

\newcommand{\beq}{\begin{equation}}
\newcommand{\eeq}{\end{equation}}
\newcommand{\ud}{\mathrm{d}}
\newcommand{\calO}{\mathcal{O}}
\newcommand{\bo}[1]{\mathbf{#1}}
\newcommand{\phm}{\phantom{-}}

\begin{document}

\title{Gravitational self-torque and spin precession in compact binaries}

\author{Sam R. Dolan}
\email{s.dolan@sheffield.ac.uk}
\affiliation{Consortium for Fundamental Physics, School of Mathematics and Statistics,
University of Sheffield, Hicks Building, Hounsfield Road, Sheffield S3 7RH, United Kingdom.}	

\author{Niels Warburton}
\affiliation{School of Mathematical Sciences and Complex \& Adaptive Systems Laboratory, University College Dublin, Belfield, Dublin 4, Ireland.}

\author{Abraham I.~Harte}
\affiliation{Max-Planck-Institut f\"ur Gravitationsphysik, Albert-Einstein-Institut, Am M\"uhlenberg 1, 14476 Golm, Germany.}

\author{Alexandre Le Tiec}
\affiliation{Maryland Center for Fundamental Physics \& Joint Space-Science Institute, 
Department of Physics, University of Maryland, College Park, MD 20742, USA.}
\affiliation{Laboratoire Univers et Th\'eories (LUTh), Observatoire de Paris, CNRS, Universit\'e Paris Diderot, 5 place Jules Janssen, 92190 Meudon, France.}

\author{Barry Wardell}
\affiliation{School of Mathematical Sciences and Complex \& Adaptive Systems Laboratory, University College Dublin, Belfield, Dublin 4, Ireland.}
\affiliation{Department of Astronomy, Cornell University, Ithaca, NY 14853, USA.}

\author{Leor Barack}
\affiliation{School of Mathematics, University of Southampton, Southampton SO17 1BJ, United Kingdom.}
	
\date{\today}

\begin{abstract}
We calculate the effect of self-interaction on the ``geodetic'' spin precession of a compact body in a strong-field orbit around a black hole. Specifically, we consider the spin precession angle $\psi$ per radian of orbital revolution for a particle carrying mass $\mu$ and spin $s \ll (G/c) \, \mu^2$ in a circular orbit around a Schwarzschild black hole of mass $M \gg \mu$. We compute $\psi$ through $\calO(\mu/M)$ in perturbation theory, i.e, including the correction $\delta\psi$ (obtained numerically) due to the torque exerted by the conservative piece of the gravitational self-field. Comparison with a post-Newtonian (PN) expression for $\delta\psi$, derived here through 3PN order, shows good agreement but also reveals strong-field features which are not captured by the latter approximation. Our results can inform semi-analytical models of the strong-field dynamics in astrophysical binaries, important for ongoing and future gravitational-wave searches.
\end{abstract}

\maketitle

\section{Introduction}
In 1916, Einstein posited three tests of his general theory of relativity. These were to exploit observations of (i) the precession of Mercury's perihelion, (ii) the deflection of light by the Sun, and (iii) the gravitational redshift. In the same year, de Sitter \cite{deS.16} outlined a fourth test based on the precession of a system's spin angular momentum. He predicted that the rotation axis of the Earth--Moon system as it moves around the Sun experiences a non-Newtonian precession of $\sim 1.9~\text{arcsec}/\text{cent}$. Lunar laser ranging has since confirmed de Sitter's prediction \cite{Sh.al.88}.

De Sitter precession, also known as the ``geodetic effect,'' is analogous to the failure of a vector on a curved surface to return to itself after being parallel-transported around a closed curve. In the weak-field, slow-motion approximation to General Relativity (GR), the spin vector $\bo{s}$ of a test gyroscope in orbit around a large, non-rotating central mass $M$ satisfies
\beq
	\frac{\ud \bo{s}}{\ud t} = \bm{\Omega}_s \times \bo{s} \, ,
	\label{eq:nonrelprec}
\eeq
where the precession frequency $\bm{\Omega}_{s} \simeq \frac{3}{2} \, \mathbf{v} \times \bm{\nabla} {\Phi}$ depends on the orbital velocity $\mathbf{v}$ and on ${\Phi} \equiv G M / (r c^2) \ll 1$. If the large central mass is itself rotating, a gyroscope will experience an additional Lense--Thirring precession \cite{LeTh.18} due to the dragging of inertial frames. Both effects were directly measured by Gravity Probe B using gyroscopes on a polar Earth orbit \cite{Ev.al.11}. For the geodetic precession, the experiment reported $\Omega_s = 6.602 \pm 0.018~\text{arcsec}/\text{yr}$, consistent with de Sitter's formula (\ref{eq:nonrelprec}).

More extreme examples of relativistic precession are found outside the Solar System. The spin of one member of the only known double-pulsar system, PSR J0737-3039, has been found to precess at a rate of $\Omega_s = 4.77 \pm 0.66~\text{deg} / \text{yr}$ \cite{Br.al2.08}. Yet, despite an orbital period of only $2.45$ hr and strong internal gravitational fields, $\Phi \sim 10^{-6}$ for this system. Opportunities to probe spin dynamics in the highly nonlinear regime of GR are emerging with the advent of x-ray spectroscopy techniques applied to accretion disks, and of gravitational-wave detector technology. 
New tests may arise in a variety of scenarios, such as the capture of strongly-bound binaries by massive black holes (an extreme analogue of de Sitter's Sun--Earth--Moon system), or the radiative inspiral of compact-object binaries. The study of spin precession in the latter scenario requires a generalization of de Sitter's formula in two respects: from the weak field to the strong field (``$\Phi \sim 1$''), and from a test gyroscope to a self-gravitating object.

In this article, we consider geodetic precession for a spinning compact body of mass $\mu$ in a circular orbit about a Schwarzschild black hole of mass $M\gg \mu$. We calculate for the first time the $\calO(\mu/M)$ shift in the geodetic precession rate caused by the back-reaction of the conservative piece of the compact body's gravitational field, which may be viewed as a ``self-torque''. Our calculation is fully general relativistic: We make no weak-field or slow-motion assumptions. The new results are accurate through $\calO(\mu/M)$ (up to a small, controllable numerical error). For simplicity, we neglect Mathisson-Papapetrou terms \cite{Ma.37, Pap} by restricting to the small-spin regime $s \ll G \mu^2 / c$, where $s$ is the spin magnitude.

Our calculation is performed using the approximations typically associated with the gravitational self-force (GSF) formalism. This is an approach to the two-body problem in GR when the mass ratio is small while curvatures and speeds may be large \cite{Ba.09,Po.al.11}. The GSF framework is complementary to post-Newtonian (PN) methods \cite{Bl.13,FuIt.07}, which are based on a weak-field, slow-motion expansion but do not require a small mass ratio. Recent work \cite{Le.al.11} has demonstrated how synergistic GSF--PN studies, augmented by non-linear simulations in numerical relativity (NR), can inform an accurate, universal model of the two-body dynamics in GR through an effective-one-body (EOB) formalism \cite{BuDa.99,BuDa.00}. There is an ongoing effort to include spin effects in this model \cite{Ta.al.13}. Here we address this problem for the first time in the GSF context. 

Hereafter, we set $G=c=1$ and use a metric signature $+2$. Latin indices $a,b,\dots$ from the beginning of the alphabet are abstract, while the letters $i,j,\dots$ refer to spatial components in a particular frame.

\section{Analysis}
\subsection{Geodetic precession}
We start by making precise the notion of geodetic precession in GR for a pointlike {\it test} particle, i.e., in the limit $\mu\to 0$ where back-reaction effects are negligible. The particle's spin $s_a$ is assumed to be nonzero, but sufficiently small so as not to affect its motion. That is, we neglect the Mathisson-Papapetrou torque \cite{Ma.37, Pap} which is a factor of $s / \mu^2$ smaller than the self-torque. All higher multipole moments are assumed to have negligible effects both on the motion and the evolution of $s_a$. Such a particle follows a timelike geodesic $\gamma$ with its spin parallel-transported along that geodesic:
\beq
	u^b \nabla_b u^a = 0 \, , \quad\quad u^b \nabla_b s_a = 0 \, .
	\label{Geodesic}
\eeq
Here, $u^a$ is the particle's four-velocity and $\nabla_b$ is the covariant derivative compatible with the spacetime metric $g_{ab}$. It follows from Eq. \eqref{Geodesic} that the magnitudes $g_{ab} u^a u^b = - 1$ and $g^{ab} s_a s_b \equiv s^2$ are conserved along $\gamma$. The product $u^a s_a$ is also conserved, consistent with the requirement that $s_a$ be spatial in the object's rest frame: $u^a s_a = 0$.

Although the spin's magnitude is conserved, its direction may precess. Consider an orthonormal triad $e^{\phantom{i}a}_i$ ($i=1,2,3$) along $\gamma$ with legs orthogonal to $u^a$. The second equation in \eqref{Geodesic} can then be written in a form similar to \eqref{eq:nonrelprec}: The spin's frame components ${(\bm{s})}_i = e^{\phantom{i}a}_i s_a$ satisfy $\ud \bm{s} / \ud \tau = \bm{\omega}_s \times \bm{s}$, where $\tau$ is the proper time along $\gamma$ and $\bm{\omega}_s$ depends on the choice of triad.

A natural class of locally-defined triads may be singled out by noting that for circular orbits, of interest here, there exists a Killing vector field $k^a$ which satisfies $k^a|_\gamma =  u^a$. It is therefore possible to choose frames that are ``comoving'' with the particle in the sense that they are Lie-dragged along $k^a$: $\mathcal{L}_k e^{\phantom{i}a}_i = k^b \nabla_b e^{\phantom{i}a}_i - e^{\phantom{i}b}_i \nabla_b k^a = 0$. For any frame within this class, it is easily shown that both $\bm{\omega}_s$ and $\bm{\omega}_s \cdot \bm{s}$ are constant along $\gamma$. Additionally, ${(\bm{\omega}_s)}_i = \frac{1}{2} \, e^{\phantom{i}a}_i \varepsilon_{abcd} \, u^b K^{cd}$, where $K_{a}{}^{b} \equiv \nabla_a k^b|_\gamma$ and $\varepsilon_{abcd}$ is the natural volume element associated with $g_{ab}$. 

We see that $\bm{s}$ undergoes a simple precession about the fixed direction of $\bm{\omega}_s$ with a proper-time frequency $\omega_s \equiv | \bm{\omega}_s|$ satisfying
\beq
    \omega_s^2  = - \frac{1}{2} \, K_{a}{}^{b} K_{b}{}^{a} .
    \label{PrecessGen}
\eeq
The frequency $\omega_s$ is manifestly independent both of the particular choice of triad within the class of Lie-dragged frames and of the angle between $\bm{s}$ and $\bm{\omega}_s$.

We now specialize the metric $g_{ab}$ to be a Schwarzschild geometry with mass $M$ and introduce Schwarzschild coordinates $(t,r,\theta,\varphi)$. We let our test body move on a circular geodesic at $\theta = \pi/2$ and $r = r_\Omega \equiv (M/\Omega^2)^{1/3}$, where $\Omega \equiv u^\varphi/u^t$ is the orbital frequency seen by a distant stationary observer. The unique Killing field which coincides with $u^a$ on $\gamma$ is explicitly $k^a = u^t \, [(\partial_t)^a + \Omega \, (\partial_{\varphi})^a]$, where $u^t = (1-3M/r_\Omega)^{-1/2}$. Direct calculation shows that $\bm{\omega}_s$ is aligned with the orbital angular momentum (thus $\bm{s}$ precesses in the same sense as the orbital motion) and has the magnitude $\omega_s = \Omega$. A convenient, intuitive measure of the spin precession effect is given by $ \psi \equiv 1 - \omega_s / u^\varphi$, the angle of spin precession per radian of orbital motion. For a test particle on a circular orbit around a Schwarzschild black hole,
\beq
	\psi(r_\Omega) = 1 - \sqrt{1 - 3M/r_\Omega} \, . 
	\label{psi0}
\eeq
%

\subsection{Self-torque effect}
Next, we endow the particle with a small mass $\mu$ and ask how the spin precession rate is modified by self-interaction at $\calO(\mu)$. The spin is assumed to be sufficiently small so as not to affect either the motion or the metric. It is a remarkable result of the GSF literature \cite{DeWh.03,Ha.12} that, subject to certain requirements on the object's compactness, the form of Eqs.\ (\ref{Geodesic}) remains valid through $\calO(\mu)$ if one merely  replaces the underlying metric with a certain smooth \textit{effective} metric $\tilde g_{ab}$. That is, the ``perturbed'' orbit $\tilde \gamma$ is a geodesic of $\tilde g_{ab}$. Similarly, the particle's spin $\tilde{s}_a$ satisfies $\tilde u^b \tilde \nabla_b \tilde s_a = 0$, where $\tilde{u}^a$ denotes the particle's four-velocity and $\tilde{\nabla}_b$ is the covariant derivative compatible with $\tilde g_{ab}$. The metric $\tilde g_{ab}$ is given by $g_{ab}+ h^R_{ab}$, where $g_{ab}$ is the background metric and the ``regular field'' $h^R_{ab}(\propto\mu)$ \cite{DeWh.03,Po.al.11} is a certain smooth solution to the vacuum Einstein equation linearized about $g_{ab}$. While $\tilde \gamma$ is a geodesic of $\tilde g_{ab}$, it can be useful to reinterpret this as an accelerated orbit with respect to ${g}_{ab}$, subject to a GSF. Likewise, $\tilde s_a$ may be said either to be parallel-transported with respect to $\tilde g_{ab}$ or to experience a ``self-torque'' with respect to $g_{ab}$. 

Here we focus on circular orbits and ``conservative'' dynamics, defined by imposing time-symmetric boundary conditions on $h^R_{ab}$. There then exists a vector field $\tilde k^a$ which is Killing with respect to $\tilde g_{ab}$ and coincides with $\tilde u^a$ on $\tilde \gamma$. Gauges may be chosen such that $\tilde k^a$ is also Killing with respect to the Schwarzschild background. Using Schwarzschild coordinates, there exist constants $\tilde u^t$ and $\tilde \Omega$ such that $\tilde k^a =\tilde u^t \, [(\partial_t)^a + \tilde\Omega \, (\partial_{\varphi})^a]$. Again, $\tilde \Omega$ represents an orbital frequency. Given $\tilde k^a$, the notion of spin precession described above for test bodies generalizes immediately. In particular, one recovers a ``tilded'' version of Eq.~\eqref{PrecessGen} with $\tilde K_{a}{}^{b} \equiv \tilde \nabla_a \tilde k^b|_{\tilde \gamma}$.

To speak of the $\calO(\mu)$ piece of the perturbed precession rate $\tilde\psi \equiv 1 - \tilde\omega_s / \tilde u^\varphi$, the perturbed worldline $\tilde\gamma$ must be associated with a fiducial background orbit $\gamma$. We let $\gamma$ be a circular geodesic of the Schwarzschild background with the same orbital frequency as $\tilde \gamma$. Hence, $\Omega = \tilde \Omega$ and $k^a = (u^t/\tilde u^t) \tilde k^a$ with $k^a|_\gamma = u^a$. One may then consider $\delta \psi \equiv \tilde{\psi} - \psi$ as a function of $\Omega$, or equivalently the ``gauge-invariant radius'' $r_\Omega = (M / \Omega^2)^{1/3}$. Using Eq.~\eqref{PrecessGen} and its tilded analog, we find
\beq
	\delta\psi = - (2 u^\varphi \omega_s)^{-1} K^{ab} \Lambda_{ab},
	\label{domegas}
\eeq
where
\beq
	\Lambda_{ab} \equiv u^c \Bigl( \nabla_{[a} h^R_{b]c} + R_{abcd} \, \delta\gamma^d \Bigr)  .
	\label{dK}
\eeq
Here, $R_{abcd}$ is the background Riemann tensor, $\delta\gamma^d$ is a deviation vector between $\tilde\gamma$ and $\gamma$, and square brackets denote antisymmetrization. The first term in Eq.~\eqref{dK} arises from the perturbation to the connection. The origin of the second term is a first-order Taylor expansion in the separation between $\tilde \gamma$ and $\gamma$. The curvature arises from this expansion via the identity $\nabla_c \nabla_a k^b = R^b_{\phantom{b}acd}k^d$, valid for all Killing fields.

For circular motion in a Schwarzschild background, Eq. \eqref{domegas} reduces to
\beq
	\delta\psi(r_\Omega) = r_\Omega \left(\Lambda^{rt}-\Lambda^{r\varphi}/\Omega\right)
	\label{domegas_simp}
\eeq
in terms of Schwarzschild coordinate components. Evaluating this requires knowledge of the deviation vector. Imposing the equation of motion gives $\delta \gamma^b = - \frac{1}{3} [ (1- 2M/r_\Omega) ( u^\varphi)^2]^{-1} a^b$ \cite{Sa.al.08}, where $a^b = \tilde{u}^a \nabla_a \tilde{u}^b$ is the ``self-acceleration.'' Directly evaluating this yields $a_b = \frac{1}{2} u^a u^c ( \nabla_b h_{ac}^R - 2 \nabla_a h_{bc}^R)$. Alternatively, $\mathcal{L}_k h_{ab}^R = 0$ may be used to write $a_b = \frac{1}{2} \nabla_b ( h^R_{ac} k^a k^c)$ \cite{De.08}. These results allow the gauge-invariant function $\delta \psi (r_\Omega)$ to be computed directly from knowledge of $\nabla_a h_{bc}^R$ at the particle's location.

\subsection{Post-Newtonian expansion}
Before presenting our numerical results for $\delta \psi$, let us derive a PN expression for this quantity, which may be used for comparison. We use an application of the Arnowitt-Deser-Misner (ADM) canonical formulation of general relativity \cite{Ar.al.62} which has been developed to describe a binary system of spinning compact objects, modeled as point particles with masses $m_A$ $(A=1,2)$ and canonical spins $\bo{s}_A(t)$ \cite{St.al2.08,StSc.09}. In the center-of-mass frame, the conservative dynamics derives from an autonomous Hamiltonian $H(\bo{r},\bo{p},\bo{s}_A)$, where $\bo{r}(t)$ and $\bo{p}(t)$ are the relative position and momentum. These satisfy the canonical algebra $\{r^i,p^j\} = \delta^{ij}$ and $\{s_A^i,s_B^j\} = \delta_{AB} \epsilon^{ijk} \! s_A^k \!$, all other Poisson brackets vanishing. To \textit{linear order} in the spins, we have the Hamiltonian
\beq
	H(\bo{r},\bo{p},\bo{s}_A) = H_\text{orb}(\bo{r},\bo{p}) + \sum_A \bm{\Omega}_{s,A}(\bo{r},\bo{p}) \cdot \bo{s}_A \, ,
	\label{H}
\eeq
where the spin-independent orbital part $H_\text{orb}(\bo{r},\bo{p})$ is known through 4PN order \cite{JaSc.98,Da.al.00,Da.al.01,JaSc.12,JaSc.13,Da.14}. The spin-orbit piece, $\sum_A \bm{\Omega}_{s,A}(\bo{r},\bo{p}) \cdot \bo{s}_A$, contributes to the equations of motion at leading 1.5PN order \cite{BaOc.79} and has been computed to a relative 2PN accuracy \cite{Da.al.08,St.al2.08,Ha.al.13}. The vectors $\bm{\Omega}_{s,A}$ are the precession frequencies of the spins with respect to coordinate time $t$. Indeed, using the Poisson algebra, one easily derives the precession equations $\ud \bo{s}_A / \ud t = \bm{\Omega}_{s,A} \times \bo{s}_A$ \cite{Da.al.08}.

For the purpose of comparison with the GSF results, we now take $m_1 = \mu$ and $m_2 = M$, set $\bo{s}_2 = \bo{0}$, and assume that $\bo{s}_1$ is sufficiently small so as not to affect the dynamics. The orbital motion then takes place in a fixed plane orthogonal to the conserved angular momentum $\bo{L}$, and we may introduce polar coordinates $(r,\varphi)$ in that plane. Using the explicit expression for the spin-orbit piece of the PN Hamiltonian \eqref{H} in ADM coordinates, $\bm{\Omega}_{s} \equiv \bm{\Omega}_{s,1}$ is found to be aligned with the orbital angular momentum: $\bm{\Omega}_{s} = (g/r^3) \, \bo{L}$, where the ``gyro-gravitomagnetic ratio'' $g$ is a function of the separation $r = |\bo{r}|$, the radial momentum $p_r = (\bo{r} / r) \cdot \bo{p}$, and the (conserved) norm $L = |\bo{L}| = p_\varphi$. To leading PN order, $g = \frac{3}{2} \frac{M}{\mu} + 2$ [recall Eq.~\eqref{eq:nonrelprec}].

For circular orbits, the Hamilton equations of motion yield $\ud r / \ud t = \partial H / \partial p_r = 0$ and $\ud p_r / \ud t = - \partial H / \partial r = 0$, where the partial derivatives are evaluated at $p_r = 0$. The first of these equations is satisfied identically, while the second yields a relationship $L(r)$. Combined with the expression for the orbital frequency $\Omega \equiv \ud \varphi / \ud t = \partial H / \partial L$ as a function of $r$ and $L$, we obtain a relation between the norm $\Omega_s = (g / r^3) \, L$ and $\Omega$. The precession rate $\psi_\text{PN} = \Omega_{s} / \Omega$ is then calculated to be
\begin{align}
	\psi&_\text{PN} = \left( \frac{3}{4} + \frac{3}{4} \Delta + \frac{\nu}{2} \right) x + \left( \frac{9}{16} + \frac{9}{16} \Delta + \frac{5}{4} \nu - \frac{5}{8} \Delta \, \nu - \frac{\nu^2}{24} \right) x^2 \nonumber \\ &+ \left( \frac{27}{32} + \frac{27}{32} \Delta + \frac{3}{16} \nu - \frac{39}{8} \Delta \, \nu - \frac{105}{32} \nu^2 + \frac{5}{32} \Delta \, \nu^2 - \frac{\nu^3}{48} \right) x^3 \nonumber \\ &+ O(x^4) \, ,
	\label{psiPN}
\end{align}
where $\nu \equiv \mu M / (M+\mu)^2$ is the symmetric mass ratio, $\Delta \equiv (M - \mu) / (M+\mu) = \sqrt{1-4\nu}$ is the reduced mass difference, and $x \equiv [(M+\mu) \Omega]^{2/3} = \calO(c^{-2})$ is the small PN parameter. Expression \eqref{psiPN} is valid for \textit{any} mass ratio. An alternative derivation was first given in Ref.~\cite{Bohe:2012mr}, based on knowledge of the 3PN near-zone metric in harmonic coordinates.

We now expand our result through $\calO(\mu^2)$, introducing the convenient inverse-radius parameter $y \equiv (M \Omega)^{2/3}=M/r_{\Omega}$ in terms of which $x = y \, (1+\mu/M)^{2/3}$. We obtain
\begin{align}
	\psi_\text{PN} &= \frac{3}{2} y + \frac{9}{8} y^2 + \frac{27}{16} y^3 + \frac{\mu}{M} \left( y^2 - 3 y^3 \right) \nonumber \\ &+ \left(\frac{\mu}{M} \right)^2 \left( - \frac{y}{3} + \frac{2}{3} y^2 + \frac{53}{8} y^3 \right) + \calO(y^4,\mu^3) \, ,
	\label{Psi_GSF}
\end{align}
where the $\calO(\mu^0)$ term is consistent with the exact test-particle result \eqref{psi0}. Interestingly, the $\calO(\mu)$ term begins at $\calO(y^2)$, so that the leading self-torque correction is suppressed at large radii by a factor $M/r_{\Omega}$ compared to the test-particle term. Notice also that, through 3PN order [$\calO(y^3$)], the self-torque \textit{increases} the precession rate beyond the usual geodetic effect. The $\calO(\mu^2)$ term in \eqref{Psi_GSF} could be compared to a future second-order perturbative calculation \cite{De.12,Gr.12,Po.12,Ha.12}.

\subsection{Numerical method and results}
A range of methods for numerically computing $h_{ab}^R$ and its derivatives are described in the GSF literature. We used two independent computational frameworks: (i) the method of Ref.\ \cite{Ak.al.13}, which is based on mode-sum regularization \cite{BaOr.00}, and (ii) the method of Ref.\ \cite{DoBa.13}, based on $m$-mode regularization \cite{Ba.al3.07}. Both computations were performed in the Lorenz gauge, apart from a minor gauge modification to the monopole sector required to bring the perturbation into an asymptotically-flat form (see discussion in Ref.~\cite{Sa.al.08}). For method (i), higher-order regularization parameters were obtained using the technique of \cite{He.al.12}. Our two sets of numerical results were found to be in agreement to within the error bars of method (ii). Method (i) provided the most accurate data, presented here.

\begin{figure}
	\includegraphics[width=8.5cm]{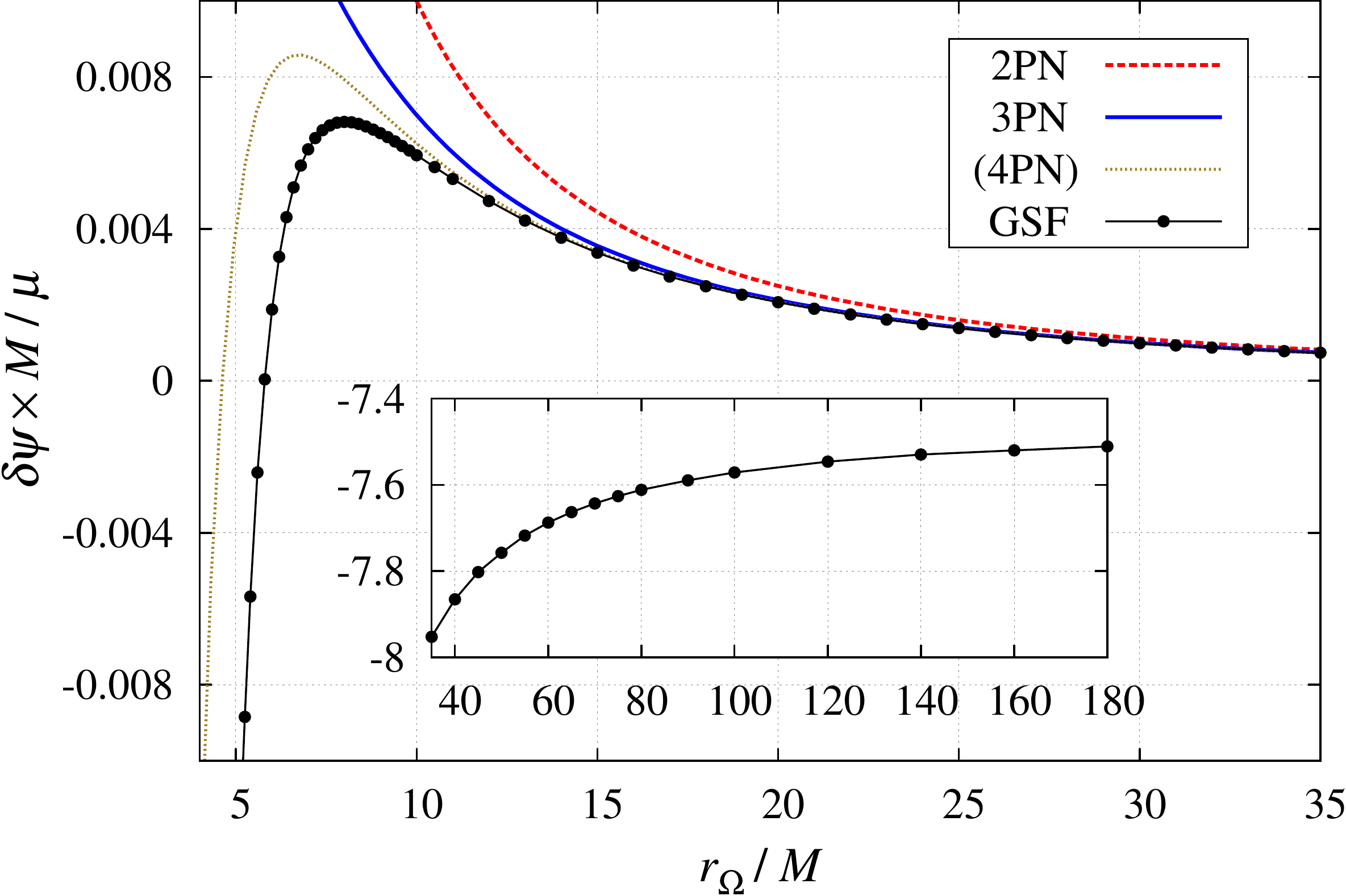}
	\caption{The $\calO(\mu)$ conservative correction to $\psi$ as a function of orbital radius $r_\Omega$. The solid black line interpolates the numerical GSF data, and the dashed red and solid blue lines show the 2PN and 3PN predictions for comparison. The inset, showing the difference between the GSF and 3PN results multiplied by $(r_{\Omega}/M)^4$, hints at the value of the yet unknown 4PN coefficient. The thin brown line displays the 4PN curve assuming that the relevant coefficient is $-15/2$.}
	\label{fig:results}
\end{figure}

Table \ref{tbl:data} and Fig.~\ref{fig:results} show numerical results for $\delta \psi(r_{\Omega})$. The data is consistent with the PN expansion \eqref{Psi_GSF}. The data supports the absence of a 1PN term proportional to $M/r_\Omega$, as well as the values of the $\calO(\mu)$ 2PN and 3PN coefficients. Moreover, the data is accurate enough to suggest that the value of the next (yet unknown) term at 4PN order is close to $-15/2$. The data shows that $\delta \psi(r_{\Omega})$ changes sign at $r_\Omega \simeq 5.8M$, below the Schwarzschild innermost stable circular orbit, becoming negative for smaller orbital radii. The maximal value of $\delta \psi(r_{\Omega})$ is reached at $r_\Omega \simeq 8.02M$. These strong-field features cannot be inferred from available PN expressions.

\begin{table}
\begin{tabular}{lll|lll}
\toprule
$r_{\Omega}/M$ & \quad\quad $\delta \psi \times M / \mu$ & & $r_{\Omega}/M$ & \quad\quad $\delta \psi \times M / \mu$ \\
\hline
4 & $-1.1669057(3)$ & $\times 10^{-1}$ & 30 & $\phm 9.900329(4)$ & $\times 10^{-4}$\\
5 & $-1.6055004(5)$ & $\times 10^{-2}$ & 35 & $\phm 7.410563(1)$ & $\times 10^{-4}$\\
6 & $\phm 1.8780855(8)$ & $\times 10^{-3}$ & 40 & $\phm 5.750523(3)$ & $\times 10^{-4}$\\
7 & $\phm 6.0923305(9)$ & $\times 10^{-3}$ & 50 & $\phm 3.747588(3)$ & $\times 10^{-4}$\\
8 & $\phm 6.8178260(4)$ & $\times 10^{-3}$ & 60 & $\phm 2.632957(4)$ & $\times 10^{-4}$\\
9 & $\phm 6.5220522(1)$ & $\times 10^{-3}$ & 70 & $\phm 1.950169(3)$ & $\times 10^{-4}$\\
10 & $\phm 5.9385649(3)$ & $\times 10^{-3}$ & 80 & $\phm 1.50204(1)$ & $\times 10^{-4}$\\
12 & $\phm 4.7347769(2)$ & $\times 10^{-3}$ & 90 & $\phm 1.19225(1)$ & $\times 10^{-4}$ \\
14 & $\phm 3.7660516(1)$ & $\times 10^{-3}$ & 100 & $\phm 9.69242(6)$ & $\times 10^{-5}$ \\
16 & $\phm 3.0367142(1)$ & $\times 10^{-3}$ & 120 & $\phm 6.76719(3)$ & $\times 10^{-5}$ \\
18 & $\phm 2.4887335(2)$ & $\times 10^{-3}$ & 140 & $\phm 4.9910(2)$ & $\times 10^{-5}$ \\
20 & $\phm 2.0715008(2)$ & $\times 10^{-3}$ & 160 & $\phm 3.8318(1)$ & $\times 10^{-5}$  \\
25 & $\phm 1.3868631(3)$ & $\times 10^{-3}$ & 180 & $\phm 3.0342(3)$ & $\times 10^{-5}$ \\
\botrule
\end{tabular}
\caption{The $\calO(\mu)$ conservative correction to $\psi$, the angle of spin precession per radian of orbital motion, for a sample of orbital radii. Parenthetical figures are estimates of the numerical error bars on the last quoted decimals.}
\label{tbl:data}
\end{table}

\section{Concluding remarks}
In this article, we have presented a first calculation of a strong-field spin precession effect beyond the geodesic approximation. This opens up a number of directions for further study.

1. Comparison with other methodologies: The functional relationship $\tilde u^t(r_\Omega)$ has been exploited extensively in recent literature to interface between the GSF, PN, NR and EOB approaches \cite{De.08,Bl.al.10,Bl.al2.10,Le.al.12,Le.al2.12,Ba.al.12,Ak.al.12}. Here we have introduced $\tilde \psi(r_\Omega)$ as a new handle on the strong-field orbital dynamics. This should allow (i) new constraints on the free parameters of the EOB model \cite{Da.10}, (ii) fresh comparisons with NR simulations using GSF coefficients with a symmetric mass ratio \cite{Le.al.11,Le.al2.12,Le.al.13}, and (iii) numerical determination of high-order PN coefficients \cite{Bl.al.10,Bl.al2.10,Ba.al.10,Le.al.12}, as we have started to demonstrate here.

2. Self-torque correction to the Lense-Thirring effect: To achieve this requires an extension of our calculation to, e.g., circular equatorial orbits on a Kerr background. Our general formulas \eqref{domegas}-\eqref{dK} still apply in this case. Methods for numerically computing $h_{ab}^R$ in Kerr are becoming available \citep{Sh.al.12}.

3. Beyond circular orbits: Although our particular formulation relies on helical symmetry, it is likely that spin precession may be defined in an orbital-average sense for more general periodic configurations such as eccentric orbits. Such an extension of our analysis would give access to further information on the two-body dynamics and enable further comparisons.

Our analysis opens up a new front in the ongoing effort to model the strong-field dynamics in binary sources of gravitational waves, which are prime targets for ground-based detectors such as Advanced LIGO \cite{LIGO} and Advanced Virgo \cite{Virgo}, and for future space-based missions such as eLISA \cite{eLISA}. We envisage that our results will help stimulate a program to accurately incorporate spin effects into models spanning the full range of binary mass ratios \cite{Ta.al.13}.

N.W.'s work was supported by the Irish Research Council, which is funded under the National Development Plan for Ireland. A.L.T. acknowledges support from NSF through Grants No.~PHY-0903631 and No.~PHY-1208881, as well as from the Maryland Center for Fundamental Physics. B.W. gratefully acknowledges support from Science Foundation Ireland under Grant No.~10/RFP/PHY2847 and from the John Templeton Foundation New Frontiers Program under Grant No.~37426 (University of Chicago) - FP050136-B (Cornell University). L.B. acknowledges funding from the European Research Council under the European Union's Seventh Framework Programme FP7/2007-2013/ERC Grant No.~304978, and additional support from STFC in the UK through Grant No.~PP/E001025/1.

\bibliography{ListeRef}

\end{document}